\documentclass[reprint,amsmath,pra,amssymb,aps,groupaddress,superscriptaddress]{revtex4-1}

\usepackage{bbm}

\usepackage{graphicx}
\usepackage{dcolumn}
\usepackage{amsmath}
\usepackage{bm}
\usepackage{color}
\usepackage{mathrsfs}
\usepackage{epstopdf}
\usepackage{amssymb}
\usepackage{subfigure}
\usepackage{enumitem}
\usepackage{multirow}
\usepackage{exscale}
\usepackage{relsize}
\usepackage{dsfont}
\usepackage{pst-node}
\usepackage{tikz-cd}

\usepackage{float}
\usepackage{mathrsfs}
\date{\today}
\usepackage{xcolor}
\usepackage[colorlinks,citecolor=blue,linkcolor=red,anchorcolor=red,urlcolor=blue]{hyperref}

\newcommand{\be}{\begin{equation}}
\newcommand{\ee}{\end{equation}}
\newcommand{\bey}{\begin{eqnarray}}
\newcommand{\eey}{\end{eqnarray}}
\newcommand{\bw}{\begin{widetext}}
\newcommand{\ew}{\end{widetext}}

\newcommand{\ba}{\begin{array}}
\newcommand{\ea}{\end{array}}
\newcommand{\bi}{\begin{itemize}}
\newcommand{\ei}{\end{itemize}}
\newcommand{\bem}{\begin{enumerate}}
\newcommand{\eem}{\end{enumerate}}

\begin{document}

\title{Temperature fluctuations in mesoscopic systems}

\author{Zhaoyu Fei} 
\affiliation{Department of Physics and Key Laboratory of Optical Field Manipulation of Zhejiang Province, Zhejiang Sci-Tech University, Hangzhou 310018, China}
\affiliation{Graduate School of China Academy of Engineering Physics,
No. 10 Xibeiwang East Road, Haidian District, Beijing, 100193, China}

\author{Yu-Han Ma}
 \email{yhma@bnu.edu.cn}
 \affiliation{Department of Physics, Beijing Normal University, Beijing, 100875, China}
 \affiliation{Graduate School of China Academy of Engineering Physics, No. 10 Xibeiwang East Road, Haidian District, Beijing, 100193, China}

\begin{abstract}
We study temperature fluctuations in  mesoscopic 
$N$-body systems undergoing non-equilibrium processes from the perspective of stochastic thermodynamics. By introducing a stochastic differential equation, we describe the evolution of the system's temperature during an isothermal process, with the noise term accounting for finite-size effects arising from random energy transfer between the system and the reservoir. Our analysis reveals that these fluctuations make the extensive quantities (in the thermodynamic limit) deviate from being extensive for consistency with the theory of equilibrium fluctuation. Moreover, we derive finite-size corrections to the Jarzynski equality, providing insights into how heat capacity influences such corrections. Also, our results indicate a possible violation of the principle of maximum work by an amount proportional to $N^{-1}$. Additionally, we examine the impact of temperature fluctuations in a finite-size quasi-static Carnot engine. We show that irreversible entropy production resulting from the temperature fluctuations of the working substance diminishes the average efficiency of the cycle as $\eta_{\rm{C}}-\left\langle \eta\right\rangle \sim N^{-1}$, highlighting the unattainability of the Carnot efficiency $\eta_{\rm{C}}$ for mesoscopic-scale heat engines even under the quasi-static limit
\end{abstract}

\maketitle

\section{Introduction}

In the field of nonequilibrium thermodynamics, stochastic thermodynamics have attracted much attention recently. Notable among its accomplishments are the fluctuation theorems, which provide a quantitative framework for understanding the statistical behavior of nonequilibrium processes and offer a generalization of the second law. They are widely found and proved in time-dependent driving processes~\cite{Sekimoto2010, Seifert2012, Klages2013, Crooks1999, Hummer2001}, nonequilibrium steady states~\cite{Evans1993, Gall1995, Maes1999, Lebowitz1999, Hatano2001}, and quantum systems~\cite{aq2000, tasaki2000, Talkner1999, Hekking2013, Esposito2009, Horowitz2013, Liu2017, Funo2018}. 

In addition to the aforementioned nonequilibrium processes, the departure of a many-body system from thermal equilibrium can be attributed to finite-size effects. In such cases, the system fluctuates around the-maximum-entropy (minimum-free-energy) state, with probabilities governed by the exponential of entropy (free energy) as per Einstein's interpretation of the reverse form of the Boltzmann entropy. Consequently, the system resides not in a full equilibrium state but rather in a quasi-equilibrium state, as postulated by the theory of equilibrium fluctuations~\cite{Einstein1910, Landau2000,Mishin2015}. 

These studies primarily focused on equilibrium fluctuations, providing limited insights into characterizing the fluctuations and thermodynamic behaviors of mesoscopic systems (whose finite size effects is significant) undergoing nonequilibrium processes. However, the lack of clarity regarding mesoscopic nonequilibrium thermodynamics has created a gap in connecting microscopic dynamics and macroscopic thermodynamics. This motivates us to establish a general framework from the perspective of stochastic thermodynamics to bridge this gap.

Given that fluctuations in a finite-size system exist at thermal equilibrium, they are also supposed to manifest in a nonequilibrium process involving such a system. In this study, we delve into the finite-size effects of a many-body system undergoing an isothermal process, viewing it through the lens of stochastic thermodynamics. 
For a system in a canonical ensemble, we propose that the finite-size effect manifests in temperature fluctuations. We formulate a stochastic differential equation (SDE) to describe the evolution of a generic system's temperature during the isothermal process, where the noise term accounts for finite-size effects stemming from random energy exchanges between the system and its reservoir (also see Refs.~\cite{Salazar2016, Chen2022} for specific systems).
The temperature fluctuation, characterized by the temperature distribution, will lead to modifications of the Jarzynski equation and the average efficiency of a quasi-static Carnot cycle at the mesoscopic scale. These modifications  result in deviations from the well-established results typically applicable to systems in the thermodynamic limit. 

This paper is organized as  follows: In Sec. II, the stochastic differential equation for the system’s temperature is derived. We further obtain the Fokker-Planck equation for the system's temperature in Sec. III. In Sec. IV, the stochastic thermodynamics in terms of the fluctuating temperature is developed and the finite-size  correction to the Jarzynski equality is obtained. As a demonstration of our theory, we study the efficiency of a finite-size heat engine in  a quasi-static Carnot cycle in Sec. V. The conclusions and discussion of the study are given in Sec. VI.  

\section{Temperature fluctuations in isothermal processes at mesoscopic scale} 

In this section, we begin by providing a brief introduction to the  stochastic Fokker-Planck equation~\cite{Fei2303}, where the noise term characterizes the fluctuation of the flux density arising from the random collisions between the system and the reservoir at the mesoscopic level. We regard the equation as the generalization of the theory of equilibrium fluctuations in  nonequilibrium processes.

Subsequently, we present the corresponding stochastic differential equation governing the system's temperature under the ergodic approximation. As a result, the stationary solution represents a quasi-equilibrium state in a canonical system, in alignment with the theory of equilibrium fluctuation (see Sec.~\ref{a3}). It is noteworthy that although the derivation of the stochastic differential equation for the system's temperature relies on the stochastic Fokker-Planck equation, we contend that our findings remain independent of the specific details concerning the system's evolution in nonequilibrium processes.  These results can be applied to study %the systems described by discrete states or other systems in nonequilibrium processes. 
discrete-state systems or other systems not described by the stochastic Fokker-Planck equation.

\subsection{Stochastic  Fokker-Planck equation}

%In the isothermal process, we assume that the working substance in the heat engine is governed by a  stochastic  Fokker-Planck equation that is a nonlinear
%equation due to the inclusion of a noise term characterizing the fluctuation of the flux density. The nonlinear equation
%determines an equilibrium state satisfying   non-Boltzmann statistics~\cite{Kaniadakis2001, Frank2005, Chavanisa2008, Fei2303}.
%The  noise term  keeps track of the finite-$N$ effect arising from random collisions between the system and the reservoir.

Let $\rho(\bm z), \bm z=(\bm x, \bm p)$, $\bm x=(x_1,\cdots,x_d)$, $\bm p=(p_1,\cdots,p_d)$ denote the one-particle phase-space distribution of the system ($d$ denotes the dimension of the system).
The stochastic Fokker-Planck equation is given by~\cite{Fei2303} (also see~\cite{Chavanis2011, Chavanis2014})
\be
\label{es1}
\frac{\partial \rho}{\partial t}=L_\text{st}\rho+ \frac{\partial }{\partial \bm p}\cdot \bm j,
\ee
with a flux density in phase space $\bm j$ originating from collisions between the system and the reservoir. It reads
\be
\label{ej}
\bm j=\gamma \bm p\rho(1+\epsilon\rho) +\gamma m k_\text{B} T_{\mathrm r}\frac{\partial \rho}{\partial \bm p}+\bm \zeta,
\ee
where $L_\text{st}=-\frac{ \bm p}{m}\cdot\frac{\partial }{\partial \bm x}+\frac{\partial U}{\partial \bm x}\cdot\frac{\partial }{\partial \bm p}$ denotes the streaming operator,
$m$ the mass of the particle, $\gamma$ the damping coefficient, $U$ the potential energy,
%(or the mean-field effective potential including long-range interactions),
$k_\text{B}$ the Boltzmann constant, and $T_{\mathrm r}$ the temperature of the reservoir. And, $\epsilon =1, -1, 0$ for (non-condensed)  bosons, fermions  and distinguishable  particles, respectively.  Here and in the following paper, we do not show the time dependence of the functions explicitly without ambiguity.

Due to the discreteness of  particle number and the randomness of   collisions between the system and the reservoir, %Eq.~(\ref{es1}) only describes the evolution of $\rho$ on average for a finite-$N$ system. Hence, we add
the noise term   characterizes the finite-size effects of the dynamics.
%Then,  we  obtain a stochastic  Fokker-Planck equation~\cite{Fei2303} (also see~\cite{Chavanis2011, Chavanis2019}):
%\be
%\label{e7}
%\frac{\partial \rho}{\partial t}=L_{\text{st}}\rho+\frac{\partial}{\partial \bm p} \cdot(\bm j+\bm \eta),
%\ee
Here, $\bm \zeta$ is a $d$-dimensional Gaussian white noise satisfying $\langle\zeta_{i }(\bm z, t)\rangle=0$, $\langle\zeta_{i}(\bm z, t)\zeta_{j }(\bm z', t')\rangle=2h^d m\gamma k_\text{B}T_{\mathrm r}\rho(\bm z, t)[1+\epsilon \rho(\bm z, t)]\delta_{ij}\delta(\bm z-\bm z')\delta(t-t')$ ($h$ denotes the Planck constant).
In the thermodynamic limit, the suppression of the noise $\zeta$  is shown in Refs.~\cite{Chavanis2011,  Fei2303}.

%Associated with the equilibrium state~(\ref{e3}), 
Equation~(\ref{es1}) is  conservative in particle number
\be
\label{een}
N =\int  \rho  \mathrm d \bm z,
\ee
where $\mathrm d\bm z=\prod_{i=1}^{d}\mathrm d x_i\mathrm d p_i/h^d$. %, and $h$ denotes the Planck constant.
The internal energy $E$ and the Boltzmann entropy $S$ of the system are respectively given by
\be
\label{eee}
E=\int \left(\frac{p^2}{2m}+U\right) \rho\mathrm d\bm z,
\ee
and
\be
\label{ees}
S=k_\text{B}\int[-\rho\ln \rho+\epsilon^{-1}(1+\epsilon \rho)\ln (1+\epsilon \rho)]\mathrm d\bm z.
\ee
Also, we define $F=E-T_\mathrm{r} S$ as the nonequilibrium free energy of the system.
%Thus, Eq.~(\ref{e3}) is the extreme distribution of $S[\rho]$ with constant $E[\rho]$ and constant $N[\rho]$.

In the absence of the noise term $\bm \zeta$, Eq.~(\ref{es1}) determines a steady state (a semiclassical equilibrium state in phase space)
\be
\label{e3}
\rho_{\text{eq}}(\bm z)=\frac{1}{\mathrm e^{\beta_{\mathrm r}\left[p^2/(2m)+U(\bm x)-\mu_{\mathrm r}\right]}-\epsilon},
\ee
where $\beta_{\rm{r}}=1/(k_\text{B} T_{\mathrm r})$ is the inverse temperature, and $\mu_{\mathrm r}$   the chemical potential, $p^{2}=\sum_{i=1}^{d}p_{i}^{2}$. In fact, the equilibrium state $\rho_{\text{eq}}$ is the  minimum point of the nonequilibrium free energy $F$ with constant  $N$. %determines the equilibrium state given in Eq. (\ref{e3}).

\subsection{Ergodic Approximation}

Let $\tau_\text{p}$ denote the characteristic time of the motion due to the potential (e. g., the oscillating period of the harmonic trap).
When $\tau_\text{p}$ is much smaller than the relaxation time $\gamma^{-1}$, the variation of $\rho$ along the equienergy surface in the phase space is relatively small.
The distribution function therefore only depends on the
phase-space variables through the energy variable $\varepsilon(z)=p^2/(2m)+U(\bm x)$.
Such an approximation is called ergodic approximation, which has been widely
used in the literatures on kinetic theory~\cite{Snoke1989, Luiten1996, Holland1997, Jaksch1997, Gardiner2017, Bijlsma2000} (in Refs.~\cite{Salazar2016, Chen2022}, it is called highly underdamped regime).

Moreover, we assume that the potential energy $U$ explicitly depends on a time-dependent parameter $\lambda(t)$, called work parameter.
The ergodic approximation requires that $\tau_\text{p}\ll \tau_\text{d} $, where $\tau_\text{d} $ is the driving time of $\lambda$. %(slow-driving condition).
Then, following the similar procedure in Ref.~\cite{Bijlsma2000} and using Eq.~(\ref{es1}), we obtain the evolution equation of the mean occupation number $\tilde{\rho}(\varepsilon)$ at the single-particle energy $\varepsilon$:
 \begin{gather}
  \begin{split}
  \label{e8}
&\frac{\partial }{\partial t}(g \tilde{\rho} )+\frac{\mathrm d\lambda}{\mathrm dt}\frac{\partial}{\partial \varepsilon}\left( \overline{\frac{\partial U}{\partial\lambda}} g\tilde{\rho} \right)\\
&=\frac{\partial }{\partial \varepsilon}\left[\frac{\gamma g\overline{p^2}}{m}\left(\tilde{\rho}+\epsilon\tilde{\rho}^{2}+k_\text{B} T_{\mathrm r}\frac{\partial \tilde{\rho}}{\partial \varepsilon}\right)+\tilde{\zeta}\right],
    \end{split}
 \end{gather}
where
\be
g(\varepsilon, t)=\int \delta\left[\varepsilon-\frac{p^2}{2m}-U(\bm x, \lambda_t)\right]\mathrm d\bm z
\ee
denotes the density of states ($\lambda_t\equiv \lambda(t)$), and
\be
\tilde{\rho}(\varepsilon, t) = g(\varepsilon,  t)^{-1}\int \delta\left[\varepsilon-\frac{p^2}{2m}-U(\bm x, \lambda_t)\right]\rho(\bm z, t)\mathrm d\bm z,
\ee
with $\tilde{\rho}(\varepsilon(z), t)=\rho(z, t)$ under the ergodic approximation. Here, we have used the abbreviation
\be
\overline{O(\bm z)}\equiv \frac{1}{g(\varepsilon, t)}\int \delta\left[\varepsilon-\frac{p^2}{2m}-U(\bm x, \lambda_t)\right]O(\bm z)\mathrm d\bm z,
\ee
and
\be
\tilde{\zeta}(\varepsilon, t)\equiv\int  \delta\left[\varepsilon-\frac{p^2}{2m}-U(\bm x, \lambda_t)\right]\frac{\bm p}{m}\cdot \bm \zeta(\bm z, t)\mathrm d\bm z
\ee
is a Gaussian white noise
satisfying $\langle \tilde{\zeta}(\varepsilon, t)\rangle=0$, $\langle\tilde{\zeta}(\varepsilon, t)\tilde{\zeta}(\varepsilon', t')\rangle=2m^{-1}\gamma k_\text{B}T_{\mathrm r}\tilde{\rho}(\varepsilon, t)[1+\epsilon \tilde{\rho}(\varepsilon, t)]g(\varepsilon, t)\overline{p^2}(\varepsilon, t)\delta(\varepsilon-\varepsilon')\delta(t-t')$. Such a noise characterizes the fluctuation of the density distribution of the system in the single-particle energy space.

\subsection{SDE for the System's Temperature}

The RHS of Eq.~(\ref{e8}) describes random collisions between the particles and the reservoir. Besides, there are also collisions among the particles. Let $\tau_\text{a}$ denote the relaxation time due to the internal collisions. We assume $\tau_\text{a}\ll \gamma^{-1}, \tau_\text{d}$ so that the system is approximately a equilibrium state during the time scales $\gamma^{-1}, \tau_\mathrm{d}$. Thus, it is characterized by a time-dependent effective temperature $T$ and a time-dependent effective chemical potential $\mu$, which is called endoreversibility~\cite{CA1975,Hoffmann2016,Ma2020finitesize,Yuan2022,Chen2022}.
Specifically,  we  have (the mean occupation number at $\varepsilon$, Eq.~\ref{e3})
\be
\label{e13}
\tilde{\rho}=\frac{1}{\mathrm e^{\beta\left(\varepsilon-\mu\right)}-\epsilon}.
\ee
Substituting Eq.~(\ref{e13}) into Eqs. ({\ref{een}}, 
 {\ref{eee}}, {\ref{ees}}), one finds
\be
\label{e14}
N=k_\text{B}T\left(\frac{\partial \ln \mathcal{Z}}{\partial \mu}\right)_\beta,
\ee
\be
\label{e15}
E=-\left(\frac{\partial \ln \mathcal{Z}}{\partial \beta}\right)_{\beta\mu},
\ee
and 
\be
\label{e15s}
S=k_\text{B} \left(\ln\mathcal Z-\beta \mu  N+\beta  E \right),
\ee
with the partition function
\be
\label{e16}
\mathcal{Z} = -\epsilon\int \ln\left[1-\epsilon \mathrm e^{\beta (\mu -\varepsilon)}\right]g \mathrm d\varepsilon.
\ee
Here, Eqs.~(\ref{e14}, \ref{e15}) determine the value of $\beta =1/(k_\text{B} T )$ and $\mu $, and $\mathcal{Z} $ is the grand canonical partition function of the system.

Equations~(\ref{e13}-\ref{e16}) connect the dynamical variables $\tilde{\rho}$ and the thermodynamic variables $T, \mu, E, S$. Accordingly, the dynamic equation Eq.~(\ref{es1}) can be represented by a thermodynamic equation.
Taking the time derivative of $E $ on both sides of Eq.~(\ref{eee})  and using Eqs.~(\ref{e8}, \ref{e13}, \ref{e16}), we obtain a stochastic differential equation for internal energy
\be
\label{e17}
\frac{\mathrm d E }{\mathrm dt}=\Lambda  \frac{\mathrm d \lambda }{\mathrm dt}+\Gamma(T_{\mathrm r}-T )+\xi ,
\ee
where
\be
\label{ela}
\Lambda =-k_\text{B}T \frac{\partial \ln \mathcal{Z} }{\partial \lambda }
\ee
denotes thermodynamic force conjugate to $\lambda $, $\Gamma\equiv\gamma dN k_\text{B}$, and
\be
\xi(t)\equiv-\int  \tilde{\zeta} (\varepsilon, t)\mathrm d\varepsilon
\ee
is a Gaussian white noise satisfying $\langle \xi(t)\rangle=0$, $\langle \xi(t) \xi (t') \rangle=2\Gamma k_\text{B} T_{\mathrm r}T(t)\delta(t-t')$.
In the derivation, we have used the identity
\be
\int \frac{
 e^{\beta \left(\varepsilon-\mu \right)}\overline{p^2}g }{[\mathrm
 e^{\beta \left(\varepsilon-\mu \right)}-\epsilon]^2}\mathrm d\varepsilon=d N m k_\text{B}T .
\ee
The first two terms on the RHS of Eq.~(\ref{e17}) correspond to the
power and the rate of heat flow respectively. The rate of heat flow satisfies Newton's law of cooling with $\Gamma$ as the cooling rate. The noise term accounts for the  random energy transfer between the system and the reservoir. For a single particle in a single-well potential, similar results  have been reported in Refs.~\cite{Salazar2016, Chen2022}. Eq.~(\ref{e17}) is thus considered as a generalization to a $N$-body system in a general potential.

As a thermodynamic equation, Eq.~(\ref{e17}) describes the evolution of the system in a nonequilibrium isothermal process, which should satisfy thermodynamic relations. To proceed, we  consider $\lambda, T, N$ as independent thermodynamic variables and do not show their dependence of functions for simplicity. We introduce $C \equiv \partial E/\partial T$  as the heat capacity with constant $\lambda$ and obtain the following thermodynamic relations from Eqs.~(\ref{e15}, \ref{e15s}, \ref{ela})
\be
\label{e23r}
\frac{\partial S}{\partial T}= \frac{C }{T},
\frac{\partial S}{\partial \lambda}= -\frac{\partial \Lambda }{\partial T},
\frac{\partial C}{\partial \lambda}=-T\frac{\partial^2 \Lambda }{\partial T^2}.
\ee

Moreover, taking the time derivative on both sides of Eq.~(\ref{e15}),  we obtain the first law by using Eqs.~(\ref{e15s}, \ref{e23r})
 \begin{gather}
  \begin{split}
\label{e21}
\frac{\mathrm d E}{\mathrm dt}=& \Lambda \frac{\mathrm d \lambda}{\mathrm dt}+T\circ\frac{\mathrm d S}{\mathrm d t}\\
=&\left(\Lambda-T\frac{\partial \Lambda}{\partial T} \right)\frac{\mathrm d \lambda}{\mathrm dt}+C\circ\frac{\mathrm d T}{\mathrm d t}
    \end{split}
 \end{gather}
where $\circ$ indicates the Stratonovich integral, which enables us to use ordinary calculus.
Due to the noise $\xi$, the Stratonovich integral and the Ito integral are related by
\be
\label{str}
C\circ\frac{\mathrm d T}{\mathrm d t}=C\frac{\mathrm d T}{\mathrm dt}+\frac{\Gamma k_\text{B}T_\mathrm{r}T}{C^2}\frac{\partial C}{\partial T}.
\ee
Comparing Eq.~(\ref{e21}) with  Eq.~(\ref{e17}) and transforming the Stratonovich integral into the  Ito integral (Eq.~\ref{str}), we finally obtain the stochastic differential equation for the system's temperature
%\begin{widetext}
 \begin{gather}
  \begin{split}
C\frac{\mathrm d T}{\mathrm dt}=& T\frac{\partial \Lambda}{\partial T} \frac{\mathrm d \lambda}{\mathrm dt}+\Gamma(T_\mathrm{r}-T)-\frac{\Gamma k_\text{B}T_\mathrm{r}T}{C^2}\frac{\partial C}{\partial T}+\xi.
\label{Eq:dT/dt}
    \end{split}
 \end{gather}
%\end{widetext}
%It is worth mentioning that although we derive the SDE for the system's temperature from a substantial way, we argue that under the above conditions, Eq.~(\ref{Eq:dT/dt}) is heavily independent of the details of how the system evolve in the isothermal process. Hence, it can be applied to study the systems with discrete states or other systems in nonequilibrium processes.

\section{\label{a3} Fokker-Planck equation for the system's temperature}

The system's temperature fluctuate due to the noise term in Eq.~(\ref{Eq:dT/dt}). The Fokker–Planck equation for  its  probability distribution $P(T, t)=\langle \delta(T-T(t))\rangle$ ($\langle\cdots\rangle$ denotes the average over the noise $\xi$) is
%Let $P(T, t)=\langle \delta(T-T(t))\rangle$  denote the probability distribution of the random temperature $T$. Using Ito calculus, it follows from Eq.~(\ref{Eq:dT/dt}) that
\begin{widetext}
 \begin{gather}
  \begin{split}
  \label{e24}
\frac{\partial P}{\partial t}=&\frac{\partial }{\partial T}\left[-\frac{T}{C}\frac{\partial \Lambda}{\partial T}\frac{\mathrm d \lambda}{\mathrm dt}P +\frac{\Gamma}{C} (T-T_\mathrm{r})P+ \frac{\Gamma k_\text{B}T_\mathrm{r}}{C}\frac{\partial}{\partial T}\left(\frac{TP}{C}\right) \right]\\
=&\frac{\partial }{\partial T}\left[-\frac{T}{C}\frac{\partial \Lambda}{\partial T}\frac{\mathrm d \lambda}{\mathrm dt}P +\frac{\Gamma  T}{C^2}P\frac{\partial }{\partial T}\left( F+ k_\text{B}T_\mathrm{r} \ln\frac{k_\text{B}TP}{C} \right)\right].
    \end{split}
 \end{gather}
\end{widetext}
%where, $F*E-T_\mathrm{r} S$ is the nonequilibrium free energy of the system, whose minimum with constant particle number determines the equilibrium state~(\ref{e3}).
Here $T\in[0, \infty)$, and we assume that $P$ quickly goes to zero when $T\to \infty$ or $T\to 0$. % so that the boundary terms in the integral by parts are negligible.
The second equality in Eq.~(\ref{e24})  shows the thermodynamic nature implied in it.
When $t\to \infty$, let $\lambda$ go to a constant. Then, Eq.~(\ref{e24}) determines a stationary solution%~\cite{Risken1996} %(notice $(\partial S/\partial T)_\lambda=C/T$)
\be
  \label{e25}
 P_\text{s}(T, \lambda)=\frac{C}{\tilde{\mathcal Z} k_\text{B}T}\mathrm e^{-\beta_r F },
\ee
where $\tilde{\mathcal Z}\equiv\int  C (k_\text{B}T)^{-1}\mathrm e^{S/k_\text{B}-\beta_r E } \mathrm d T$ denotes the generalized partition function of the system~\cite{Fei2303}. % (we do not show the explicit $\lambda$ dependence of $S$, $E$,  $C$, and $\tilde{\mathcal Z}$  for simplicity).
In the absence of the factor $C/ (\tilde{\mathcal Z}k_\text{B} T)$, $P_\text{s}$ is actually the quasi-equilibrium state in a canonical system according to the theory of equilibrium fluctuation~\cite{Mishin2015, Einstein1910, Landau2000} and satisfies the large deviation principle.

Similar to the equilibrium free energy in statistical mechanics, we define
\be
\label{eqgf}
\mathcal F\equiv-k_\text{B}T_\mathrm{r}\ln\tilde{\mathcal Z}
\ee
as the generalized free energy of the system. Then, we have the relation $\mathcal F=\mathcal E-T_\mathrm{r}\mathcal{S}$, where
\be
\mathcal E \equiv\int EP_\mathrm{s} \mathrm dT=- \frac{\partial \ln\tilde{\mathcal Z }}{\partial \beta_\mathrm{r}}
\ee
is the mean internal energy of the system at the quasi-equilibrium state, and
\be
\label{esr}
\mathcal S \equiv\int \left(S-k_\text{B}\ln\frac{k_\text{B}TP_\mathrm{s}}{C}\right) P_\mathrm{s} \mathrm dT =k_\text{B}\beta_\mathrm{r}(\mathcal E-\mathcal F )
%=k_\text{B}(\ln\tilde{\mathcal Z_\mathrm{r}}+\beta_\mathrm{r}E_\mathrm{r})
\ee
is the mean entropy of the system at the quasi-equilibrium state. Here, the mean entropy is a sum of the mean Boltzmann entropy (the first term in the integral) and the contribution from the distribution of the system's temperature (the second term in the integral), the latter of which is consistent with the perspective of information theory~\cite{Jaynes1957, Jaynes1983}. Also, we confirm the fundamental relation in thermodynamics
\be
\mathrm{d} \mathcal E=T_\mathrm{r}\mathrm d\mathcal{S}+ \tilde{\Lambda}\mathrm d \lambda,
\ee
and
\be
\mathrm{d} \mathcal F=-\mathcal{S}\mathrm{d}T_\mathrm{r}+ \tilde{\Lambda}\mathrm d \lambda,
\ee
where $\tilde{\Lambda}\equiv-k_\text{B}T_\mathrm{r} \partial \ln\tilde{\mathcal{Z}}/\partial \lambda$ denotes the generalized thermodynamic force conjugate to $\lambda$ at the quasi-equilibrium state.
We want to emphasize here that these generalized quantities $\mathcal F, \mathcal E, \mathcal S$, which are extensive in the thermodynamic limit, are no longer extensive due to the finite-size effect of the system.

In the thermodynamic limit, %and for $T_r\gg \sqrt{C_\lambda/k_\text{B}}$,
we apply the Gaussian approximation of Eq.~(\ref{e25}) as (central limit theorem)
\be
 P_\text{s}(T, \lambda)\simeq\sqrt{\frac{C_\mathrm{r} }{2\pi k_\text{B}T_\mathrm{r}^2}}\exp\left[-\frac{C_\mathrm{r} (T-T_\mathrm{r})^2}{2k_\text{B}T_\mathrm{r}^2}\right],
 \label{eq:P(T)}
\ee
where $C_\mathrm{r}\equiv C |_{T=T_\mathrm{r}}$.
%\be
%\frac{T^*_r}{T_r}\simeq 1+\frac{C_r^2}{k_{\text{B}}}\left( \frac{\partial }{\partial T_r}\frac{T_r}{C_r}\right)_\lambda
%\ee
%is the average temperature of the working substance.
%Due to the noise, there is a $N^{-1}$ correction between $T^*_r$ and $T_r$ (also see Ref.~\cite{Mishin2015}).
Eq.~(\ref{eq:P(T)}) approximates the system's  temperature fluctuation at the equilibrium state. Its mean value $T_\mathrm{r}$ and variance $k_\text{B}T_\mathrm{r}^2/C_\mathrm{r} $ are  consistent with the theory of equilibrium fluctuation~\cite{Mishin2015, Einstein1910, Landau2000}.
Then, substituting Eq.~(\ref{eq:P(T)}) into Eq.~(\ref{eqgf}),  we obtain
\be
\label{eps}
\mathcal F =F_\mathrm{r}-\frac{k_\text{B} T_\mathrm{r}}{2}\ln \frac{2\pi C_\mathrm{r}}{ k_\text{B}}+O\left(\frac{1}{N}\right),
\ee
where $F_\mathrm{r}\equiv F|_{T=T_\mathrm{r}}$. The first term on the RHS is the equilibrium free energy of the system at temperature $T_\mathrm{r}$ and
the second term on the RHS is  the finite-size correction to it.
%As an anolog to the partition function in the statistical mechanics, we define $\mathcal{F}_\mathrm{r}=-k_\text{B}T_r\ln\tilde{\mathcal Z_\mathrm{r}}$ as the general partition function of the quasi-equilibrium state.

%Or, we use the Gaussian approximation of Eq.~(\ref{e25}) in terms of internal energy as
%\be
% P_\text{s}(T)\simeq\sqrt{\frac{1 }{2\pi k_\text{B}C_\mathrm{r}T_\mathrm{r}^2}}\exp\left[-\frac{(E-E_\mathrm{r})^2}{2k_\text{B}C_\mathrm{r} T_\mathrm{r}^2}\right],
% \label{eqes}
%\ee
%where $E_\mathrm{r} =E |_{T=T_\mathrm{r}}$.

%Here, we extend the range of $T$ from $[0,\infty)$ to $(-\infty,\infty)$ since the probability of the event $T<0$ is extremely small in the thermodynamic limit.

\section{Stochastic thermodynamics}

In the spirit of stochastic thermodynamics~\cite{Sekimoto2010, Seifert2012, Klages2013}, we are going to define stochastic thermodynamic quantities  corresponding to Eqs.~(\ref{e17}, \ref{e21}, \ref{e24}) in this section. Firstly, a trajectory of the system's temperature is defined as $T_{[0,\tau]}:=\{T(t)|t\in[0,\tau]\}$.
According to Eqs.~(\ref{e17}, \ref{e21}), the stochastic work $w[T_{[0, t]}]$ and the stochastic heat $q[T_{[0,t]}]$ are respectively given by
\be
\label{ework}
w[T_{[0, \tau]}]=\int_{0}^{\tau} \Lambda  \mathrm d \lambda,
\ee
and
 \begin{gather}
  \begin{split}
q[T_{[0, \tau]}]=&\int_{0}^{\tau} T \circ \mathrm d S \\
=&\int_{0}^{\tau} \left(-T\frac{\partial \Lambda}{\partial T}   \mathrm d \lambda +C \circ \mathrm d T \right)\\
=&\int_{0}^{\tau}\left[\Gamma\left( T_{\mathrm r}-T \right)+\xi\right]\mathrm dt.
%=&-\int_{0}^{\tau}\Gamma\left(T- T_{\mathrm r} \right)\mathrm dt.
    \end{split}
 \end{gather}
Thus, we have the conservation law of energy 
\be
E(\tau) -E(0)=w[T_{[0, \tau]}]+q[T_{[0, \tau]}]
\ee
at the mesoscopic level.

Corresponding to Eq.~(\ref{e24}),
the stochastic entropy $s(t)$  %the entropy change of the reservoir $s_\text{r}[T_{[0, t]}]$,
and stochastic free energy $f(t)$
are respectively given by (also see Ref.~\cite{Fei2303})
 \begin{gather}
  \begin{split}
  \label{e29}
s(t)=&S(t)-k_\text{B}\ln\frac{k_\text{B}T(t)P(T(t), t)}{C(t)}\\
=&k_\text{B}\left[\beta_\mathrm{r}E(t)-\beta_\mathrm{r}\mathcal{F}(t)- \ln\frac{P (T(t) , t)}{P_\text{s}(T(t), \lambda(t))}\right],
    \end{split}
 \end{gather}
 and
 \begin{gather}
  \begin{split}
\label{e30}
f(t)=&E(t)-T_\mathrm{r}s(t)\\
=&F(t)+k_\text{B}T_\mathrm{r}\ln\frac{k_\text{B}T(t)P(T(t), t)}{C(t)}\\
=& \mathcal{F} (t)
+k_\text{B} T_\mathrm{r}\ln\frac{P(T(t), t)}{P_s(T(t), \lambda(t))}.
    \end{split}
 \end{gather}
Here, $P(T, t)$ is the solution  of Eq.~(\ref{e24}), $S$ ($F$) %in Eq.~(\ref{e29}) (Eq.~ \ref{e30} )
denotes the Boltzmann entropy (nonequilibrium free energy) of the system,  and the term $-k_\text{B}\ln [k_\text{B}T P C ^{-1}] $ ($k_\text{B}T_\mathrm{r}\ln [k_\text{B}T P C ^{-1}] $) denotes the finite-size correction to  $S$ ($F$) from the distribution of the system's temperature.
The term $k_\text{B}\beta_\mathrm{r}(E - \mathcal{F} )$  corresponds to the mean entropy at the quasi-equilibrium state $\mathcal S$ in Eq.~(\ref{esr}). And the term $-k_\text{B} \ln (P /P_\text{s}  )$ (after taking the average over $P$)  corresponds to the relative entropy which measures how far the temperature distribution is from the quasi-equilibrium state. Consequently, we   have  $\mathcal{E} =\langle E\rangle|_{P=P_\mathrm{s}}$,  $\mathcal{S} =\langle s\rangle|_{P=P_\mathrm{s}}$, and  $\mathcal{F} =\langle f\rangle|_{P=P_\mathrm{s}}$.
It is worth mentioning that the stochastic free energy $f$, the difference between which and the generalized free energy $\mathcal F$ measures how far the system departures from the quasi-equilibrium state, has not been reported in previous papers (but for Ref.~\cite{Fei2303}).

Moreover,  the stochastic total entropy production $s_{\text{p}}[T_{[0,\tau]}]$ reads
 \begin{gather}
  \begin{split}
\label{e31}
s_{\text{p}}[T_{[0, \tau]}]=& s(\tau)-s(0)  +s_\text{r}[T_{[0, \tau]}]
    \end{split}
 \end{gather}
where    $s_\text{r}[T_{[0, \tau]}]=-q[T_{[0, \tau]}]/T_\mathrm{r}$ is the stochastic entropy change of the reservoir.

Then, we prove the fluctuation theorems based these stochastic quantities. According to Eq.~(\ref{Eq:dT/dt}), the   probability distribution of the trajectory $T_{[0, \tau]}$ conditioned with a fixed initial temperature $T_0\equiv T(0)$ reads~\cite{Lau2007, Cugliandolo2017, Aron2016}
\be
\label{e322}
  P[T_{[0, \tau]}|T_0]=  \mathrm e^{- \tilde{\mathcal S}[T_{[0, \tau]}]},
\ee
where the integral measure is $\mathcal DT\equiv\prod_{i=1}^N\mathrm{d}T_i\sqrt{C_i^{2}/(2\pi T^*_i\Delta t )}$, with the Stratonovich discretization $0=t_0<t_1<\cdots<t_{N-1}<t_N=\tau$, $\Delta t\equiv t_i-t_{i-1}$, $T_i\equiv T(t_i)$, $T_i^*\equiv(T_i+T_{i-1})/2$,   $\lambda_i\equiv\lambda(t_i)$, $\lambda_i^*\equiv(\lambda_i+\lambda_{i-1})/2$, $C_i \equiv C|_{T=T_i^*,\lambda=\lambda_i^*}$.
%\be
%\mathcal DT=\prod_{i=1}^N\mathrm{d}T_i\sqrt{ \frac{C(T^*_i)^2}{2\pi\Delta t T^*_i}}
%\ee
Here, the action $\tilde{\mathcal S}$ as a generalized Onsager-Machlup functional is given by
\begin{widetext}
 \begin{gather}
  \begin{split}
    \label{e32}
\tilde{\mathcal S}[T_{[0, \tau]}]=&\frac{1}{4\Gamma k_\text{B} T_\mathrm{r}}\int_{0}^{\tau}\left[C \frac{\mathrm d T}{\mathrm dt}-T \frac{\partial \Lambda }{\partial T } \frac{\mathrm d \lambda }{\mathrm dt}-\Gamma(T_\mathrm{r}-T )-\frac{\Gamma k_\text{B} T_\mathrm{r}T }{C }\frac{\partial }{\partial T }\ln\frac{C }{T } \right]^2 \frac{\mathrm dt}{T }\\
&+\frac{1}{2}\int_{0}^{\tau}  \frac{\partial }{\partial T }\left[\frac{T }{C }\frac{\partial \Lambda }{\partial T } \frac{\mathrm d \lambda }{\mathrm dt}+\frac{\Gamma}{C }(T_\mathrm{r}-T )-\frac{\Gamma k_\text{B} T_\mathrm{r}}{2C ^2} \right] \mathrm dt.
    \end{split}
 \end{gather}
\end{widetext}
In Eq.~(\ref{e32}), we also have chosen the Stratonovich discretization. Such a choice makes sure that the time reversal of $P[T_{[0, \tau]}|T_0]$ is also under the Stratonovich discretization~\cite{Cugliandolo2017, Aron2016}.

To proceed, let  $ P^\dag[T^{ \dag}_{[0, \tau]}|T^{ \dag}_0] $ denote the conditional probability distribution of the  reverse  trajectory  $T_{[0, \tau]}^{ \dag}:=\{T(\tau-t)|t\in[0, \tau]\} $  with another fixed initial temperature  $T^{ \dag}_0\equiv T^{ \dag}(0)$  and a   reverse protocol  $\lambda^\dag (t):=\lambda(\tau-t)$ (the overscript $\dag$ indicates the reverse trajectory).
It follows from Eqs.~(\ref{e322}, \ref{e32}) that
 \begin{gather}
  \begin{split}
  \label{e321}
 P^\dag[T^\dag_{[0, \tau]}|T^\dag_0]= & \mathrm e^{-\tilde{ \mathcal S}[T^\dag_{[0, \tau]}]}\\
=&  \left.\mathrm e^{-\tilde{ \mathcal S}[T_{[0, \tau]}]}\right|_{\{ \frac{\mathrm d T}{\mathrm d t}, \frac{\mathrm d \lambda} {\mathrm d t}\}\to \{-\frac{\mathrm d T}{\mathrm d t}, -\frac{\mathrm d \lambda} {\mathrm d t}\}},
    \end{split}
 \end{gather}
and thus the detailed fluctuation theorem is obtained as
 \begin{gather}
  \begin{split}
\ln\frac{P[T_{[0, \tau]}|T_0]}{P^\dag[T^{ \dag}_{[0, \tau]}|T^{ \dag}_0]}=\ln\frac{P (T_\tau, \tau)}{P(T_0, 0)}+\frac{s_\text{p}[T_{[0, \tau]}]}{k_\text{B}},
    \end{split}
 \end{gather}
where $T_\tau\equiv T(\tau)$. By adding an arbitrary normalized distribution   at the  initial time of the  reverse process $P'(T^\dag_0, 0)$ and noticing $\mathcal{D}T=\mathcal{D}T^\dag$, we obtain the integral fluctuation theorem as
\be
\left\langle  \frac{P'(T^\dag_0, 0)}{P(T_\tau, \tau)} \mathrm e^{-\frac{s_\text{p}} {k_\text{B}}}\right\rangle=1.
\ee
Such an equality is formally consistent with the integral fluctuation theorems in previous studies~\cite{Seifert2012, Klages2013}. For a choice of  $P'(T^\dag_0, 0)=P(T_\tau, \tau)$, we obtain the integral fluctuation theorem for total entropy production~\cite{Seifert2012, Klages2013, Fei2303}
\be
\left\langle    e^{-s_\text{p} /k_\text{B}}\right\rangle=1.
\ee
As a corollary, the second law $\left\langle s_\text{p} \right\rangle\geq 0$ follows from the fluctuation theorem   by using Jensen's inequality.

When both $P(T_0, 0), P'(T^\dag_0, 0)$  are stationary solutions of the Fokker-Planck equation (quansi-equilibrium state in Eq.~(\ref{e25})), i. e., $P(T_0, 0)=P_\text{s}(T_0, \lambda_0) , P'(T^\dag_0, 0)=P_\text{s}(T_\tau, \lambda_\tau) $, we obtain the generalized Jarzynski equality~\cite{Fei2303} 
\be
\left\langle   \mathrm  e^{-\beta_\mathrm{r} w }\right\rangle=\mathrm e^{-\beta_\mathrm{r}\Delta\mathcal{F} },
\ee
and the generalized principle of maximum work $\langle w \rangle \geq \Delta\mathcal{F} $ by using Jensen's inequality~\cite{Fei2303}, where $\Delta A\equiv A(\tau)-A(0)$ for some time-dependent function $A$. %The term $S$ in .
In the thermodynamic limit,  it follows from  Eq.~(\ref{eps}) that
\be
\left\langle \mathrm  e^{-\beta_\mathrm{r} w }\right\rangle= e^{-\beta_\mathrm{r} \Delta F_\mathrm{r}  }\sqrt{\frac{C_{\mathrm{r}}(\lambda_\tau)}{C_{\mathrm{r}}(\lambda_0)}}+O\left(\frac{1}{N}\right),
\ee
and
\be
\label{emw}
\left\langle  w \right\rangle\geq \Delta F_\mathrm{r}-\frac{k_\text{B} T_\mathrm{r} }{2}\ln\left[\frac{C_{\mathrm{r}}(\lambda_\tau)}{C_{\mathrm{r}}(\lambda_0)}\right]+O\left(\frac{1}{N}\right).
\ee
Here, the ratio of the heat capacity is a finite-size correction to the Jarzynski equality and the principle of maximum work. Eq.~(\ref{emw}) indicates that when $C_{\mathrm{r}}(\lambda_\tau)>C_{\mathrm{r}}(\lambda_0)$, a possible violation of the principle of maximum work by an amount on the order of  $N^{-1}$ is possible.
We further define the following quantity to characterize such a correction
\begin{equation}
\phi\equiv\left|\frac{\ln\left[C_{\mathrm{r}}(\lambda_\tau)/C_{\mathrm{r}}(\lambda_0)\right]}{2\beta_\mathrm{r} \Delta F_\mathrm{r} }\right|. \label{phi}
\end{equation}
%$\phi$ can be specifically obtained with certain system.

\begin{figure} 

\includegraphics[width=8.5cm]{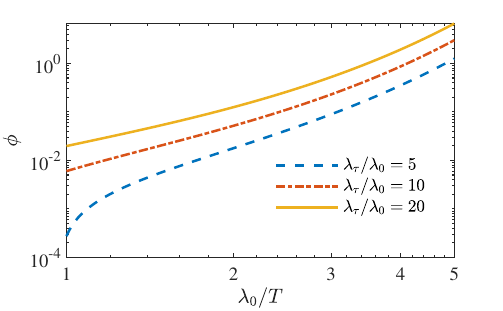}\caption{$\phi$ as a function of $T$ for different $\lambda_{\tau}/\lambda_{0}$ }
\label{fig:Je}
\end{figure}

For example, if the system is specified as $N=1000$ two-level particles with energy spacing $\lambda$ (See Appendix \ref{TLS} for details), we plot $\phi$ as a function of  temperature $T_\mathrm{r}$ in Fig. \ref{fig:Je} for different $\lambda_\tau/\lambda_0$ . In this figure, we see that $\phi$ significantly increases  with the temperature decreases. It is hence possible in principle to observe the finite-size correction to Jarzynski equality in some experimental platforms. In addition, for systems with quantum phase transitions \cite{quan2006,ma2017}, the dependence of heat capacity on parameters near the critical point is remarkable. In such cases, the finite size correction will be particularly important for the Jarzynski equality .

\section{Fluctuating Carnot cycle}

\begin{figure}[ht]
\includegraphics[width=1\columnwidth]{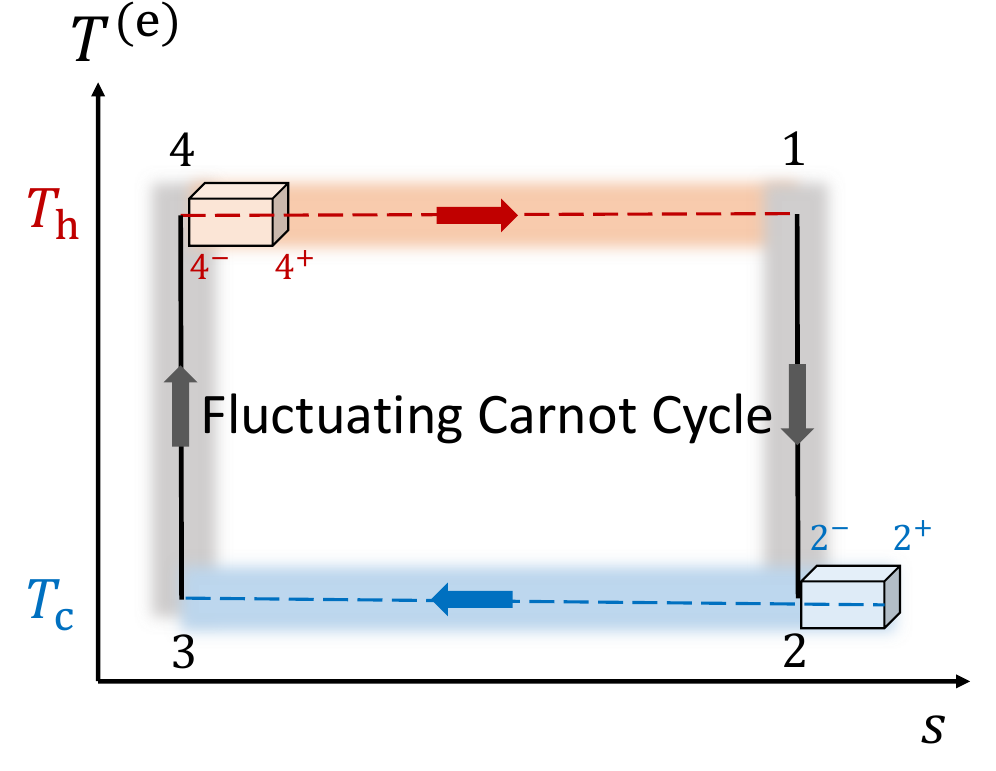}
    \caption{The Carnot cycle in the temperature-entropy diagram. The shaded regions represent the temperature fluctuation due to the finite-size effect of the working substance}
    \label{fig: Cycles}
\end{figure}

As an application of  our theory, we study a fluctuating Carnot cycle with finite-size working substance by using Eq.~(\ref{Eq:dT/dt}).
As illustrated in Fig. \ref{fig: Cycles}, the Carnot cycle consists of four processes:  $1\to 2$, adiabatic compression ($\gamma=0$);  $2\to 3$, isothermal expansion  (hot reservoir's temperature $T_\mathrm{h}$);  $3\to 4$, adiabatic expansion ($\gamma=0$);  $4\to 1$, isothermal compression (cold reservoir's temperature $T_\mathrm{c}$).  Let $T_n, n=1,\cdots,4$ denote the corresponding temperature  of the working substance at state $n$. Then, we have $\langle T_1\rangle=\langle T_4\rangle=T_\text{c}$ and $\langle T_2\rangle=\langle T_3\rangle=T_\text{h}$. Let $\lambda_n, E_n, C_n, S_n, s_n $, $n=1,\cdots,4$ denote the corresponding work parameter, internal energy, heat capacity,  Boltzmann entropy, stochastic entropy of the working substance respectively. By using Eq.~(\ref{e24}) with $\gamma=0$, it is straightforward to prove that in the adiabatic processes, both the average Boltzmann entropy and stochastic entropy are constants, i. e.,
\be
\label{eqes1}
 \langle S_1\rangle=\langle S_2\rangle,\quad \langle S_3\rangle=\langle S_4\rangle,
\ee
and
\be
\label{eqes2}
 \langle s_1\rangle=\langle s_2\rangle,\quad \langle s_3\rangle=\langle s_4\rangle.
\ee

Then, let $P_n, P_{\text{s}n}, \mathcal{F}_n$, $n=1,\cdots,4$ denote the corresponding temperature distribution, quasi-equilibrium state, and generalized free energy of the working substance respectively. In the quasi-static limit, $\mathrm d\lambda/\mathrm dt\to 0$ and the working substance is a quasi-equilibrium state with $T_\mathrm{r}=T_\mathrm{h}$ ($T_\mathrm{r}=T_\mathrm{c}$) all the time in the isothermal expansion (compression) process according to Eq.~(\ref{e24}). Meanwhile according to Appendix~\ref{TLS2}, the working substance at the end of the adiabatic processes is generally not a quasi-equilibrium anymore. That is to say $P_1=P_{\text{s}1}, P_3=P_{\text{s}3}$, and $P_2\neq P_{\text{s}2}, P_4\neq P_{\text{s}4}$. Such a result reflects the fact that from state $2^-\to 2^+$ ($4^-\to 4^+$), the working substance have quickly thermalized during a vanishing small time scale $\gamma^{-1}$, where the average temperature of the working substance remain the same but the variance of the temperature of the working substance changes. Consequently, no work is done during the vanishing small time and the finite irreversible entropy production occurs due to the contribution from the relative entropy (see Eqs.~\ref{e29}-\ref{e31}), which is a finite-size effect and firstly reported in Ref.~\cite{quan2014} to our best knowledge. 

In Fig.~\ref{fig: Cycles}, we illustrate the finite-size effects of the fluctuating Carnot cycle in terms of the temperature and entropy of the working substance. The end to end solid and dashed lines represent the mean values of them, while the shaded regions signify the fluctuations attributable to finite-size effects. We accentuate the irreversible entropy production depicted within the two cuboids, which diminishes the average efficiency of the Carnot cycle.

%In the quasistatic limit, $\mathrm d\lambda/\mathrm dt\to 0$ and the working substance is a quasiequilibrium state (Eq.~\ref{eq:P(T)}) with $T_\mathrm{r}=T_\mathrm{h}$ ($T_\mathrm{r}=T_\mathrm{c}$) all the time in the isothermal expansion compression) process according to Eq.~(\ref{e24}).

In the two isothermal processes, the variances of the input work both vanish (see the example in Ref.~\cite{Speck2004}). Therefore, the input work in the two isothermal processes   is a constant, i. e., $ \mathcal{F}_3-\mathcal{F}_2+\mathcal{F}_1-\mathcal{F}_4$ corresponding to the generalized principle of maximum work. In the two adiabatic processes, there is no heat transfer and the input work   is equal to the internal energy change according to the conservation law of energy, i. e., $ E_2-E_1+E_4-E_3$. Thus using Eq.~(\ref{e29}), the total input work reads
 \begin{gather}
  \begin{split}
  \label{eqw}
w_\text{in}=&E_2-E_1+E_4-E_3+\mathcal{F}_3-\mathcal{F}_2+\mathcal{F}_1-\mathcal{F}_4\\
=&T_\mathrm{h}\left(s_2-s_3+k_\text{B}\ln\frac{P_2}{P_{\text{s}2}}\right)+T_\mathrm{c}\left( s_4-s_1+k_\text{B}\ln\frac{P_4}{P_{\text{s}4}}\right).
    \end{split}
 \end{gather}
Accordingly, the absorbed heat from the hot reservoir reads
\be
  \label{eqq}
q_{\text h}=T_\mathrm{h}\left(s_3-s_2-k_\text{B}\ln\frac{P_2}{P_{\text{s}2}}\right),
\ee
where, $s_1, s_2$ are independent of $s_3, s_4$ due to the thermalization in the two isothermal processes. Note that the connections between $s_1, s_2$ or $s_3, s_4$ satisfy the energy-conservation equation in the adiabatic process (Eq.~\ref{Eq:dT/dt} with $\gamma=0$).
It follows from Eq.~(\ref{e31}) that the entropy production of the cycle is (also see Refs.~\cite{sinit2011, campisi2014})
 \begin{gather}
  \begin{split}
  \label{eesp}
s_\text{p}=&\frac{q_{\text h}+w_\text{in}}{T_\text{c}}-\frac{q_{\text h}}{T_\text{h}} \\
=&s_4-s_1-s_3+s_2+k_\text{B}\ln\frac{P_4}{P_{\text{s}4}}+k_\text{B}\ln\frac{P_2}{P_{\text{s}2}}.
    \end{split}
 \end{gather}
% \begin{gather}
%  \begin{split}
%  \label{e67}
%\frac{\mathrm d T }{\mathrm dt}=&\frac{T }{C }\frac{\partial \Lambda }{\partial T } \frac{\mathrm d \lambda}{\mathrm dt}.
%    \end{split}
% \end{gather}

To study the efficiency of the cycle, we adopt the definition of the stochastic efficiency in Ref.~\cite{Fei2022}
\be
\label{eta}
\eta=-\frac{w_\text{in}}{\langle q_\text{h}\rangle},
\ee
which is called the scaled fluctuating efficiency. The moments of the efficiency always exist, and its mean value is equal to the conventional efficiency of a cycle. Using Eqs.~(\ref{eqes2}-\ref{eqq}), the average efficiency of the cycle is
 \begin{gather}
  \begin{split}
  \label{eqeta}
\langle \eta \rangle=&1-\frac{T_\mathrm{c}\left[\Delta \overline{s} +D(P_4||P_{\text{s}4}) \right]}{T_\mathrm{h}\left[\Delta \overline{s} -D(P_2||P_{\text{s}2})\right]}\\
=&\eta_\text{C}-(1-\eta_\text{C})\frac{\langle s_\mathrm{p}\rangle}{\Delta \overline{s}}+O\left(\frac{1}{N^2}\right).
    \end{split}
 \end{gather}
Here, $\eta_\text{C}\equiv 1-T_\mathrm{c}/T_\mathrm{h}$ is the Carnot efficiency, $\Delta \overline{s}\equiv \langle s_3\rangle- \langle s_2\rangle=\langle s_4\rangle- \langle s_1\rangle$ is the average entropy change of the working substance in the isothermal expansion process,
\be
D(P||P_{\text s})\equiv \int P\ln\frac{P}{P_\text{s}}\mathrm dT,
\ee
is the relative entropy, and $\langle s_\mathrm{p}\rangle= D(P_2||P_{\text{s} 2})+D(P_4||P_{\text{s}4})$ following from Eqs.~(\ref{eqes2}, \ref{eesp}) is the total average entropy production of the cycle according to Eq.~(\ref{eesp}). Since $\Delta \overline{s}>0$ and $\langle s_\mathrm{p}\rangle\geq 0$, we conclude that the irreversible entropy production due to the temperature fluctuation of the working substance diminishes the average efficiency of the cycle. Consequently, even in the quasi-static limit, the Carnot efficiency remains unattainable. Such an equation is also shown in Ref.~\cite{quan2014}. %Moreover, the positive work requires $\langle \eta \rangle>0$ and we have by using Eq.~(\ref{eqeta})
%\be
%\frac{T_\mathrm{h}-T_\mathrm{c}}{T_\mathrm{c}}>\frac{\langle s_\mathrm{p}\rangle}{\Delta \overline{s}}.
%\ee
%In another word, there is a lower bound on the temperature difference between the hot and cold reservoirs for the working of a small-size heat engine ($ \langle s_\mathrm{p}\rangle/\Delta \overline{s} \sim 1/N$).

In the thermodynamic limit, we are only concerned about the mean value $\langle A \rangle$ and the variance $\sigma^2_A=\left\langle (A-\left\langle A\right\rangle)^2\right\rangle$ for some $T$-dependent function $A$.
As a result, the temperature distribution of the working substance is approximately a Gaussian distribution, i. e., $T_n\sim \mathcal{N}(\langle T_n\rangle, \sigma^2_{T_n})$
%\be
%P_n\simeq \frac{1}{\sqrt{2\pi \sigma^2_n}}\mathrm e^{- \frac{(T-\langle T_n\rangle)^2}{2\sigma^2_n}},
%\ee
for $n=1,\cdots,4$.
Therefore, we find
\be
\label{eds}
\Delta \overline{s}=S_{3\text{h} }-S_{2\text{h} }+O\left(\frac{k_\text{B}}{N}\right)=S_{4\text{c} }-S_{1\text{c} }+O\left(\frac{k_\text{B}}{N}\right),
\ee
and 
\be
\label{e55}
\langle s_\mathrm{p}\rangle=\frac{1}{2}\left[\ln(\kappa\kappa')+\frac{1}{\kappa}+\frac{1}{\kappa'}-2\right],
\ee
where $S_{n\text{c}(\text{h})}\equiv S_n|_{T=T_{\text{c}(\text{h})}}$ for $n=1,\cdots,4$, $\kappa\equiv C(T_\mathrm{c}, \lambda_2)/C(T_\mathrm{h}, \lambda_1)$, and $\kappa'\equiv C(T_\mathrm{h}, \lambda_4)/C(T_\mathrm{c}, \lambda_3)$
(see Appendix~\ref{TLS2}).

As an example, we specific the working substance as the $N$-particles system studied in Ref.~\cite{quan2014}, with the following Hamiltonian
\be
H=\sum_{i=1}^{dN}\left(\frac{p_i^2}{2m}+a\left|\frac{x_i}{L}\right|^\lambda \right)+V,
\ee
where $m$ denotes the mass of the $N$ particles, $a$ the characteristic energy of the system, $L$ the characteristic length of the system, $\lambda$ the work parameter, and $V$ the interactions among these particles (which can be ignored in comparison with the kinetic and potential energy but is strong enough to make the particles be ergodic). Taking use of the internal energy of the system at the equilibrium state (Eq.~B15 in Ref.~\cite{quan2014}), we obtain the total average entropy production of the cycle from Eq.~(\ref{e55}) as
 \begin{gather}
  \begin{split}
&\langle s_\mathrm{p}\rangle=\\
& \frac{1}{2}\ln\left[\frac{\lambda_1(\lambda_2+2)\lambda_3(\lambda_4+2)}{(\lambda_1+2)\lambda_2(\lambda_3+2)\lambda_4 }\right]+\frac{(\lambda_2-\lambda_1)}{\lambda_1(\lambda_2+2)}+\frac{(\lambda_4-\lambda_3)}{\lambda_3(\lambda_4+2)},
    \end{split}
 \end{gather}
with $\kappa=\lambda_1(\lambda_2+2)/[(\lambda_1+2)\lambda_2]$, and $\kappa'=\lambda_3(\lambda_4+2)/[(\lambda_3+2)\lambda_4]$.
Such a result is consistent with the leading order  of the expression of the relative entropy shown in Eq.~(7) of Ref.~\cite{quan2014} in the large-$N$ limit, while the latter was previously obtained through complicated calculation.  Such a consistence indicates that our formalism is universal and is able to exactly capture the entropy production of the cycles due to the finite-$N$ effects of the working substance.

Substituting Eqs. (\ref{eds}, \ref{e55}) into Eq. (\ref{eqeta}), the average efficiency is obtained as
 %\begin{gather}, 
  %\begin{split}
\be
\langle \eta \rangle=\eta_\text{C}-(1-\eta_\text{C})\frac{ \kappa\kappa'\ln(\kappa\kappa')+\kappa + \kappa'-2\kappa\kappa' }{2\kappa\kappa'(S_{3\text{h}}-S_{2\text{h}})}+O\left(\frac{1}{N^2}\right).
\ee
    %\end{split}
 %\end{gather}
In particular, for a constant heat capacity of the working substance  (such as ideal gas),  $\kappa=\kappa'=1$ and the Carnot efficiency  is recovered $\langle \eta \rangle=\eta_\text{C}$ to the order of  $N^{-1}$.

Furthermore, we consider the variance of the efficiency $\sigma_{\eta}^2$. It follows from Eq.~(\ref{eta}) that $\sigma_{\eta}^2=\sigma^2_{w_\text{in}}/\langle q_\mathrm{h}\rangle^2$.
Using the linear relation between the initial and final internal energy in the adiabatic process (Appendix~\ref{TLS2}), we  obtain the correlation functions
\be
\label{ee75}
\langle E_1E_2\rangle-\langle E_1 \rangle\langle  E_2\rangle=k_\text{B} T_\text{c}T_\text{h}C_{1\text{c}}+O\left( k_\text{B}^2T^2 \right),
\ee
\be
\label{ee76}
\langle E_3E_4\rangle-\langle E_3 \rangle\langle  E_4\rangle=k_\text{B} T_\text{c}T_\text{h}C_{3\text{h}}+O\left( k_\text{B}^2T^2 \right).
\ee
Combining Eqs.~(\ref{eqw}, \ref{eqq}, \ref{eds}), one finds
 \begin{gather}
  \begin{split}
\sigma_{\eta}^2=&\frac{ \sigma^2_{E_2-E_1}+\sigma^2_{E_4-E_3} }{\langle q_\mathrm{h}\rangle^2}\\
=&\frac{k_\text{B} (T_\text{h}-T_\text{c})^2(C_{1\text{c}}+C_{3\text{h}})}{T_\mathrm{h}^2(S_{3\text{h} }-S_{2\text{h} })^2}+O\left(\frac{1}{N^2}\right).
    \end{split}
 \end{gather}
It is worth mentioning here that such a result also appeared  in Ref.~\cite{denzler2021} for the spectra of the working substance with scale property, while our result is not limited to this case. That is to say that our theory is independent of the details of the working substance and is thus universal.

\section{Conclusion and Outlook}

In this paper, we have studied the temperature fluctuations of a finite-size system. Initially, we drive a stochastic differential equation to describe the evolution of the system's temperature during a isothermal process. The noise term in the equation accounts for finite-size effects resulting from random energy exchanges between the system and its reservoir. Consequently, the system's stationary state represents a quasi-equilibrium state according to the theory of equilibrium fluctuation, in which the generalized thermodynamic quantities (which are extensive in the thermodynamic limit) deviate from being extensive due to finite-size effects.

Furthermore, we develop stochastic thermodynamics based on the temperature fluctuations of the system and substantiate the fluctuation theorems. The obtained results provide finite-size corrections to the Jarzynski equality, which is quantified by the square root of the ratio of the system's heat capacities at the final and initial stages of the driving. Additionally, we observe a breach of the principle of maximum work by an amount on the order of $N^{-1}$.

To demonstrate the impact of temperature fluctuations at the mesoscopic level in typical thermodynamic processes, we analyze the efficiency of a finite-size heat engine operating in a quasi-static Carnot cycle. Our findings reveal that even under the quasi-static limit, the Carnot efficiency $\eta_C$ remains unattainable due to the irreversible entropy production arising from temperature fluctuations of the working substance. For some specific models, our general results of the mean value and variance of the efficiency align with previous findings in Refs.~\cite{quan2014} and \cite{denzler2021}, respectively. 

In closing, several theoretical predictions made in this work can potentially be verified on a mature experimental platform \cite{chui1992}. As possible extensions of our current results, the finite-time performance \cite{Ma20181,Chen2022,Fei2022} and optimization \cite{CA1975,Ma20182,Ma20201/t} of the proposed fluctuating Carnot cycle are worth further investigation. Moreover, considering both the temperature fluctuations of the working substance and the finiteness of the heat reservoirs \cite{Ondrechen1981,Izumida2014,Ma2020finitesize,Yuan2022,Ma2023AJP}, finding the power-efficiency trade-off relation \cite{Holubec2015,Shiraishi2016,Ma20181}  and relevant optimizations of the heat engine present another challenging task with  implications for mesoscopic heat engines.

\begin{acknowledgments}
        Y. H. Ma thanks the National Natural Science Foundation of China for support under grant No. 12305037.  
\end{acknowledgments}

\clearpage
\onecolumngrid

\appendix

\section{Heat capacity of a two-level system \label{TLS}}

We consider an ensemble of non-interacting two level systems, such
as $N$ free spin-1/2 particles. The excited state and ground state
of the $i$th subsystem are denoted as $\left\rfloor e\right\rangle _{i}$
and $\left\rfloor g\right\rangle _{i}$, respectively, and the Hamiltonian
of the system reads
\begin{equation}
H=\sum_{i=1}^{N}\lambda\left\rfloor e\right\rangle _{ii}\left\langle e\right|,
\end{equation}
where $\lambda$ is the energy spacing of the two states and the ground
state energy is set to zero. When this system is at the thermal equilibrium
state with inverse temperature $\beta$, the population in the excited and ground
state are
\begin{equation}
p_{e}=\frac{\mathrm e^{-\beta\lambda}}{1+\mathrm e^{-\beta\lambda}},
\end{equation}
\begin{equation}
p_{g}=\frac{1}{1+\mathrm e^{-\beta\lambda}},
\end{equation}
with the partition function $Z(\lambda)=1+e^{-\beta\lambda}$,  and the internal energy
\begin{equation}
U=Np_{e}\lambda=\frac{N\lambda}{1+\mathrm e^{-\beta\lambda}}.
\end{equation}
Then, the heat capacity of the system is
\begin{equation}
C=\frac{\partial U}{\partial T}=\frac{N\beta^{2}\lambda^{2}\mathrm e^{\beta\lambda}}{\left(1+\mathrm e^{-\beta\lambda}\right)^{2}},\label{eq:capacity}
\end{equation}
where the $\lambda$-dependent heat capacity  contributes to the
finite-size correction of the Jarzynski equality.

By using Eq. (\ref{eq:capacity}) and noting that
\begin{equation}
\Delta F =-N\beta^{-1}\ln\left[\frac{Z(\lambda_{\tau})}{Z(\lambda_{0})}\right]=N\beta^{-1}\ln\left(\frac{1+\mathrm e^{-\beta\lambda_0}}{1+\mathrm e^{-\beta\lambda_\tau}}\right),
\end{equation}
the quality $\phi$ defined in Eq. (\ref{phi}) is specifically obtained as
\begin{equation}
\phi=
\frac{1}{2N}\left|\frac{2\ln\left[\frac{\lambda_{\tau}\left(1+\mathrm e^{-\beta\lambda_0}\right)}{\lambda_{0}\left(1+\mathrm e^{-\beta\lambda_\tau}\right)}\right]+ \beta\left(\lambda_{\tau}-\lambda_{0}\right) }{\ln\left(\frac{1+\mathrm e^{-\beta\lambda_0}}{1+\mathrm e^{-\beta\lambda_\tau}}\right)}\right|.
\end{equation}
%Obviously, $\phi$ decays with $N$ linearly and in the lower temperature
%regime, it seems that $\phi$ will have a relative large value.

\section{Solution of the system's temperature in the adiabatic process  \label{TLS2}}

%In the thermodynamic limit, we are only concerned about the average temperature $T_\mathrm{e}=\left\langle T\right\rangle$ and the variance of the temperature $\sigma_\mathrm{e}^2=\left\langle (T-\left\langle T\right\rangle)^2\right\rangle$.
Taking the average on the both sides of Eq.~(\ref{Eq:dT/dt}) with $\gamma=0$ (adiabatic process), we have
 \begin{gather}
  \begin{split}
  \label{ete}
\frac{\mathrm d  T_\mathrm{e}  }{\mathrm dt}=&\left\langle \frac{T}{C }\frac{\partial \Lambda }{\partial T} \right\rangle\frac{\mathrm d \lambda }{\mathrm dt}\\
&=\left[ \frac{T_\mathrm{e}}{C_{\mathrm{e}}} \frac{\partial \Lambda_{\mathrm{e}}}{\partial T_\mathrm{e}}+\frac{\sigma_T^2}{2}\frac{\partial^2}{\partial T_\mathrm{e}^2}\left(\frac{T_\mathrm{e}}{C_{\mathrm{e}}} \frac{\partial \Lambda_{\mathrm{e}}}{\partial T }\right)\right]\frac{\mathrm d \lambda }{\mathrm dt} +O\left(\frac{T }{N^2 \tau}\right),
    \end{split}
 \end{gather}
where, $T_\mathrm{e}\equiv\langle T\rangle$, $\Lambda_\mathrm{e}\equiv\Lambda|_{T=T_\mathrm{e}}$, and $C_\mathrm{e}\equiv C|_{T=T_\mathrm{e}}$.
Since the adiabatic process connects two isothermal processes, let $T_a$  ($T_b$) denote the temperature of the reservoir at the initial (final) time $t=0$ ($t=\tau$) and we have the following boundary conditions
\be
T_\mathrm{e}(0)=T_a,\quad T_\mathrm{e}(\tau)=T_b.
\ee
%Such conditions constrains the time-dependence of $\lambda$ in the adiabatic processes.

Let $\delta T(t)\equiv T(t)-T_\mathrm{e}(t)$. In the thermodynamic limit, we have $\delta T(t)\sim T_\mathrm{e}N^{-1/2}\to 0$. Therefore, the linearization of Eq.~(\ref{Eq:dT/dt}) with $\gamma=0$ around $T_\mathrm{e}(t)$ reads
 \begin{gather}
  \begin{split}
  \label{eleq}
\frac{\mathrm d \delta T}{\mathrm dt}=&\delta T \frac{\partial }{\partial T_\mathrm{e}}\left( \frac{T_\mathrm{e}}{C_{\mathrm{e}}} \frac{\partial \Lambda_{\mathrm{e}}}{\partial T_\mathrm{e}}\right)\frac{\mathrm d \lambda}{\mathrm dt} +O\left(\frac{T }{N \tau}\right)\\
=&\delta T\frac{\mathrm d}{\mathrm dt}\ln\frac{T_\mathrm{e}}{C_{\mathrm{e}}}+O\left(\frac{T }{N \tau}\right),
    \end{split}
 \end{gather}
where we have used Eq.~(\ref{ete}) in the second equality. The solution of Eq.~(\ref{eleq}) is
\be
\label{eleq2}
 \delta T(\tau)=\delta T(0) \frac{T_b C_a( \lambda_0)}{T_a C_b(  \lambda_\tau)} .
\ee
That is to say, $T(\tau)$ linearly depends on $T(0)$. Moreover, the standard deviation reads
\be
\label{adiabatic equation}
\sigma_{T}(\tau)=\sigma_{T}(0)\frac{T_b C_a( \lambda_0)}{T_a C_b(  \lambda_\tau)}.
\ee
Eqs.~(\ref{eleq2}, \ref{adiabatic equation}) serve as the equation for the temperature fluctuation of the system in the adiabatic process.

For an adiabatic process with an initial quasi-equilibrium state
\begin{equation}
\label{ep0}
P(T, 0)=\sqrt{\frac{ C_a( \lambda_0)}{2\pi k_{\mathrm{B}}T_a^{2} }}\exp\left[-\frac{C_a( \lambda_0) (T-T_a)^{2}}{2k_{\mathrm{B}}T_a^{2}}\right],
\end{equation}
 $\sigma_{\mathrm{e}}^{2}(0)= k_{\mathrm{B}}T_{\mathrm{e}}^{2}(0)/C_{\mathrm{e}}(0) $.
According to Eq.~(\ref{adiabatic equation}), the corresponding temperature distribution at the end of the adiabatic process during time $\tau$ is obtained as
 \begin{gather}
  \begin{split}
\label{ep1}
P (T, \tau) %& =\sqrt{\frac{1}{2\pi\sigma_{\mathrm{e}}^{2}(t)}}\exp\left[-\frac{(T_{\mathrm{e}}(t)-T)^{2}}{2\sigma_{\mathrm{e}}^{2}(t)}\right]\\
 & =\sqrt{\frac{\kappa C_b(\lambda_\tau)}{2\pi k_{\mathrm{B}}T_b^2}}\exp\left[-\frac{\kappa C_b(\lambda_\tau)(T-T_b)^{2}}{2k_{\mathrm{B}}T_b^{2} }\right],
    \end{split}
 \end{gather}
where $\kappa \equiv C_b(\lambda_\tau)/C_a(\lambda_0)$ is the ratio of the heat capacity at the final and initial times. Hence for $\kappa \neq 1$, the work substance at time $t$ is not at an quasi-equilibrium state any more.
%Besides, because the noise vanishes in the adiabatic process, the Ito integral and the Stratonovich integral are equivalent. Therefore, the average entropy change of the working substance (Eq.~\ref{e29}) does not change, i. e., $\langle s\rangle |^t_0=\langle S\rangle |^t_0-k_\text{B}\langle \ln [k_\text{B}TP(T) C^{-1}]\rangle |^t_0=0$, where we have used
%\be
%\langle S\rangle |^t_0=\int_{0}^{\tau} \left\langle\frac{\mathrm d E-\Lambda   \mathrm d \lambda  }{T }\right\rangle=0
%\ee
%due to Eq.~(\ref{e67}), and
%\be
%\langle \ln [k_\text{B}TP  C^{-1}]\rangle |^\tau_0=0
%\ee
%using Eqs.~(\ref{adiabatic equation}, \ref{ep0}, \ref{ep1}). and considering $y$ is a conserved quantity.

\end{document}